\def\year{2022}\relax
\documentclass[letterpaper]{article} 
\usepackage{aaai22}  
\usepackage{times}  
\usepackage{helvet}  
\usepackage{courier}  
\usepackage[hyphens]{url}  
\usepackage{graphicx} 
\usepackage{natbib}  
\usepackage{caption} 
\DeclareCaptionStyle{ruled}{labelfont=normalfont,labelsep=colon,strut=off} 
\usepackage{algorithm}
\usepackage{algorithmic}
\usepackage{amsmath}
\usepackage{amsfonts}
\usepackage{subcaption}
\usepackage[dvipsnames]{xcolor}
\usepackage{tabularx}
\usepackage{booktabs}
\usepackage{multirow}
\usepackage{newfloat}
\usepackage{listings}
\floatstyle{ruled}
\newfloat{listing}{tb}{lst}{}
\floatname{listing}{Listing}
\title{
No Task Left Behind: Multi-Task Learning of \\Knowledge Tracing and Option Tracing for Better Student Assessment
}
\author{
    Suyeong An\equalcontrib, 
    Junghoon Kim\equalcontrib, 
    Minsam Kim\equalcontrib, 
    Juneyoung Park\thanks{This author is the corresponding author.}\\
}
\affiliations{
    Riiid AI Research\\

    521, Teheran-ro, Gangnam-gu,\\ 
    Seoul, Republic of Korea\\
    \{suyeong.an, junghoon.kim, minsam.kim, juneyoung.park\}@riiid.co
}

\begin{document}
\maketitle

\begin{abstract}
Student assessment is one of the most fundamental tasks in the field of AI Education (AIEd). 
One of the most common approach to student assessment is Knowledge Tracing (KT), which evaluates a student's knowledge state by predicting whether the student will answer a given question correctly or not. 
However, in the context of multiple choice (polytomous) questions, conventional KT approaches are limited in that they only consider the binary (dichotomous) correctness label (i.e., correct or incorrect), and disregard the specific option chosen by the student. 
Meanwhile, Option Tracing (OT) attempts to model a student by predicting which option they will choose for a given question, but overlooks the correctness information. 
In this paper, we propose Dichotomous-Polytomous Multi-Task Learning (DP-MTL), a multi-task learning framework that combines KT and OT for more precise student assessment.
In particular, we show that the KT objective acts as a regularization term for OT in the DP-MTL framework, and propose an appropriate architecture for applying our method on top of existing deep learning-based KT models. 
We experimentally confirm that DP-MTL significantly improves both KT and OT performances, and also benefits downstream tasks such as Score Prediction (SP). 
\end{abstract}

\section{1. Introduction}
The field of AI Education (AIEd) is concerned with developing AI systems that facilitate human learning, and has the potential to provide personalized education to a wider audience at an affordable cost. Student assessment, the process of evaluating a student's knowledge level, is one of the most fundamental tasks in AIEd. 
Proper student assessment can then be used for many downstream educational tasks, such as score prediction (SP) \citep{su2018exercise, yin2019quesnet, choi2021assessment} and personalized content recommendation \citep{chen2005personalized, wang2008content, wang2016structured, ai2019concept}.


Knowledge Tracing (KT)  \citep{corbett1994knowledge} models a student's knowledge state by predicting whether the student will answer a given question correctly or not. Due to the method's simplicity and domain-agnosticity, 
KT has been extensively used for student assessment in AIEd \citep{choi2020towards, piech2015deep}. 
However, for educational contents that involve multiple choice (polytomous) questions, KT only considers the binary (dichotomous) label of correctness, and does not consider the student's option of choice. 
KT thus fails to distinguish between students who answered a given question incorrectly, when in reality, one student's answer might have been closer to the correct one than the other student's.

Option Tracing (OT) \citep{ghosh2021option, thissen1984response} is an approach that explicitly models the student's response, i.e., option choice, to a multiple choice question. 
While OT takes into account the student's option choice information, it does not consider the student's correctness.
As a result, OT may fail to trace the student's knowledge state properly. 
This motivates a multi-task learning scheme that leverages both correctness labels and option labels.

In this paper, we propose Dichotomous-Polytomous Multi-Task Learning (DP-MTL), where the model learns to predict both the student's correctness and option choice for a given question. 
This way, DP-MTL can track the student's knowledge state at a more granular level. 
In our experiments, we demonstrate that DP-MTL indeed improves KT, OT, and SP performances.  
We expect that DP-MTL will enable a more accurate student representation learning, which would, in turn, benefit many other downstream educational tasks. 

The main contributions of our paper are as follows:
\begin{itemize}
\setlength\itemsep{0.6em}
\item We introduce DP-MTL and show that, in this framework, the KT objective acts as a regularization term for OT.  
\item Additionally, we propose an architecture design for combining KT and OT on top of existing KT models.
\item We experimentally confirm that our method significantly improves KT, OT, and SP performances when applied on top of three popular KT models, based on two different datasets. 
\end{itemize}

To the best of our knowledge, this is the first work that combines KT and OT for better student assessment.

\newpage
\section{2. Related Works}
\label{sec:2}

\subsection{Knowledge Tracing}
\label{subsection:KTCTSP}
Knowledge Tracing (KT) is a student assessment task that models a student's knowledge state by predicting whether a student will answer a given question correctly or not. Dichotomous Item Response Theory (D-IRT) models ~\cite{kingston1982feasibility, way1990investigation, chen2005personalized} predict the student's answer correctness using extracted user and item parameters, where a user parameter and an item parameter each corresponds to the user's skill and the question's difficulty, respectively. 

Collaborative filtering (CF) method, which is equivalent to the multi-dimensional D-IRT method (except for the absence of sigmoid function) \citep{vie2019knowledge}, models each user (item) as a user (item) vector, instead of a scalar value. 
Though CF was originally developed for recommendation systems \citep{koren2009matrix}, CF has been extensively used for KT \citep{khosravi2017riple, vie2019knowledge}.
Recently, Neural Matrix Factorization (NMF) methods proposed to replace the conventional dot product used in CF with neural network computations \citep{he2017neural,ijcai2017-447}.

Sequential KT approaches model the student's learning trajectory, as opposed to modeling the interactions at an atomic level as done in D-IRT and CF.
Bayesian Knowledge Tracing (BKT) \citep{corbett1994knowledge} is the original KT method, which traces the student's knowledge state based on hidden Markov model.
Recently, a lot of research effort went into applying various deep learning architectures for sequential KT, including RNN-based models \citep{piech2015deep, minn2020bkt}, Dynamic Key-Value Memory Networks (DKVMN) \citep{zhang2017dynamic}, and transformer-based models \citep{pandey2019self, choi2020towards}. 


All the aforementioned methods consider KT exclusively, and do not perform OT. 
We apply our proposed method DP-MTL on top of three popular KT methods, namely, (1) D-IRT, (2) collaborative filtering, and (3) LSTM-based KT, and demonstrate that DP-MTL consistently improves the KT performance across all models considered. 

\subsection{Option Tracing}
Option Tracing (OT) \citep{ghosh2021option} is a student assessment task that traces a student's knowledge state by predicting the student's exact answer choice given a multiple choice question. 
Polytomous IRT (P-IRT), in an analogous manner to D-IRT, predicts the student's option choice using the extracted user parameters and item option parameters.
Recently, \citet{ghosh2021option} proposed to perform OT based on modified deep KT models for a more accurate student assessment. 
However, (1) they did not consider a multi-task learning setup that simultaneously performs both KT and OT; and (2) their architecture cannot take account of subtle details when performing OT (e.g., permuting the options (A,B,C) to (B,A,C) will not change the prediction from $(p_{A}, p_{B}, p_{C})$ to $(p_{B}, p_{A}, p_{C})$). 
We not only consider the multi-task learning of KT and OT, but also propose an appropriate deep learning architecture accordingly.

\subsection{Score Prediction}
Student score prediction (SP) is another important student assessment task we consider in this work \citep{sweeney2015next, iqbal2017machine, loh2020data}. Prior SP methods rely on collaborative filtering \citep{elbadrawy2016domain, sweeney2016next}, and regression models \citep{morsy2017cumulative, ren2019grade}. Other recent methods utilize KT algorithms, and address SP as a downstream task \cite{liu2019ekt, choi2021assessment}. We also treat SP as a downstream task, and show that DP-MTL leads to improved SP performance. 

\subsection{Multi-Task Learning}
Multi-Task Learning (MTL) \citep{caruana1997multitask} is a machine learning approach that trains a model to perform several related tasks simultaneously. MTL posits that training signals from a particular domain help form inductive bias for other related tasks. In this work, we demonstrate that KT and OT are an example of such related tasks that, when performed simultaneously, are mutually beneficial.

\section{3. Methodology}
In this section, we propose Dichotomous-Polytomous Multi-Task Learning (DP-MTL) that learns to perform both KT and OT simultaneously. 
In particular, we show that in the DP-MTL framework, the multi-task learning objective has an explicable interpretation: the KT loss acts as a regularization term for OT. 
Also, we propose an architecture design necessary for applying DP-MTL on top of existing KT models.



\begin{figure*}[t!]
    \centering
    \includegraphics[width=2.1\columnwidth]{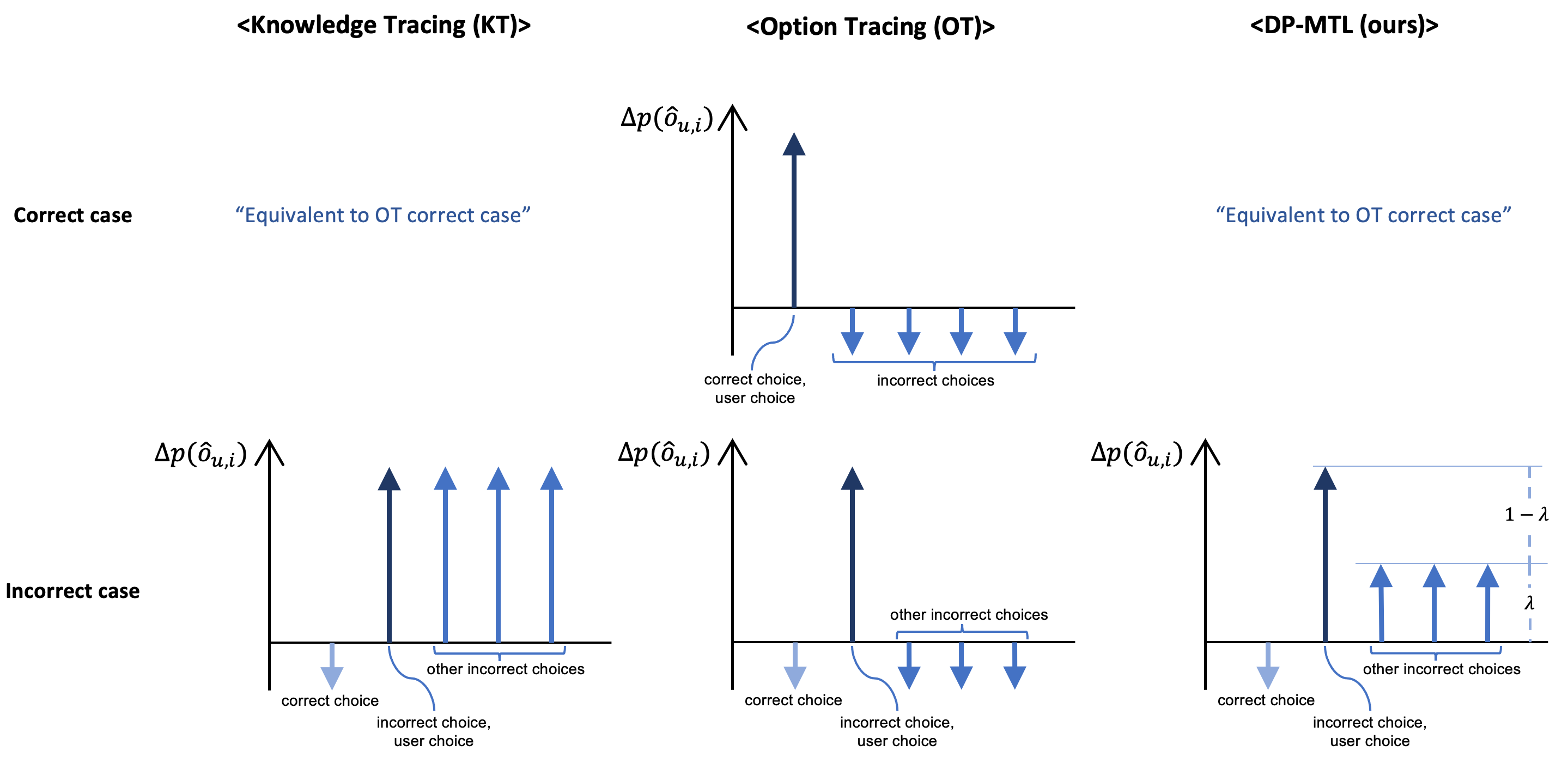}
    \caption{Illustration of DP-MTL. When the student chooses the correct answer, KT, OT, and DP-MTL are equivalent (Upper). However, when the student chooses the incorrect answer, DP-MTL interpolates between KT and OT. (Lower).}
    \label{fig:learning_dp_mtl}
\end{figure*}

\subsection{Notations}
This section introduces the notations that will be used throughout this paper.

\paragraph{Users}
Within the total population of \textit{n} students, each student \textit{u}, $1 \leq \textit{u} \leq \textit{n}$, has a \textit{d} dimensional user parameter $\boldsymbol{\theta}_{u} \in \mathbb{R}^{d}$. Each dimension should model a user's skill level in a particular knowledge component (KC).
\paragraph{Choices/Options}
The possible choice set for each question $i$ with $j$ total multiple choice options is denoted by $\mathcal{O}_i = \{o_{i}^{1}, o_{i}^{2}, \dots, o_{i}^{j}$\}. In other words, user $u$'s choice for given item $i$ is $o_{u,i} \in \mathcal{O}_{i}$. The correct choice $o_i^*$ for given item(question) $i$ must always be within the set of choices ($o_i^* \in \mathcal{O}_{i}$). 
\paragraph{Items}
For a standardized multiple choice question examination with a total of \textit{m} questions, each question \textit{i} ($1 \leq \textit{i} \leq \textit{m}$) with choice \textit{k} ($1 \leq \textit{k} \leq \textit{j}$) has \textit{d} dimensional item parameters per each choice, $\textbf{a}_{i,k} =(\textit{a}_{i,k}^{1}, a_{i,k}^{2}, \dots, a_{i,k}^{d})  \in \mathbb{R}^{d}$. For simplicity, we denote $\textbf{a}_{i}$ to be the set of item parameters with each choice $\textbf{a}_{i} =(\textbf{a}_{i,1}, \textbf{a}_{i,2}, \dots, \textbf{a}_{i,j}$) 

\subsection{DP-MTL Training}
\paragraph{Dichotomous Option Correctness (D)}
The conventional dichotomous model is trained by minimizing the negative log likelihood of observing the interactions that consists of the user, item, and the pair's corresponding \textit{correctness}. This is equivalent to maximizing the conditional probability of the user responding correctly/incorrectly to the item, based on the student interaction data. In other words,


\begin{equation}
\begin{split}
L_{D}(y_{u,i};\boldsymbol{\theta}_{u},\textbf{a}_{i}) =            
 &\; y_{u,i}\textnormal{log}P(\hat{y}_{u,i}=1 |\theta_{u},\textbf{a}_{i})
 \\ 
 &\; +(1-y_{u,i})\textnormal{log}P(\hat{y}_{u,i}=0 |\boldsymbol{\theta}_{u},\textbf{a}_{i})
\end{split}
\label{eqn:dirt}
\end{equation}
is minimized, where $\hat{y}_{u,i}$ is the prediction for $u$ getting the question $i$ correctly, and $y_{u,i}$ is the correctness label included in $\{0, 1\}$.

\paragraph{Polytomous Option Choice (P)}
Training of a polytomous model is done by minimizing the negative log likelihood $L_{P}(\theta_u, $\textbf{a}$_{i})$ of a user $u$ responding to a question $i$ with choice $o_{u,i}$:
\begin{equation}
\begin{split}
    L_{P}(o_{u,i};\boldsymbol{\theta}_u, \textbf{a}_{i}) = \textnormal{log}P(\hat{o}_{u,i}=o_{u,i}|\boldsymbol{\theta}_u, \textbf{a}_{i}).
\end{split}
\end{equation}
Here, $\hat{o}_{u,i}$ is the predicted categorical variable that represents the option choice of student $u$ for the given question $i$, and $o_{u,i}$ is the option label.


Substituting $P(\hat{y}_{u,i}=0|\boldsymbol{\theta}_{u},$\textbf{a}$_{i})$ from Equation \ref{eqn:dirt} with the sum of probabilities of incorrect choices $\sum\nolimits_{o_i^j\not=o_i^*}P(\hat{o}_{u,i}=o_i^j|\boldsymbol{\theta}_{u},$\textbf{a}$_{i})$, $L_{D}$ becomes

\begin{equation}
\begin{split}{}
    L_{D} = & y_{u,i}\textnormal{log}P(\hat{o}_{u,i}=o_i^*|\boldsymbol{\theta}_{u},\textbf{a}_{i}) 
    \\ &+(1-y_{u,i})\textnormal{log}[\sum\nolimits_{o_i^j\not=o_i^*}P(\hat{o}_{u,i}=o_i^j|\boldsymbol{\theta}_{u},\textbf{a}_{i})]
\end{split}
\end{equation}

\paragraph{DP-MTL}
DP-Multi Task Learning (DP-MTL) is a combined version of option correctness (D) and option choice (P) with a ratio of $\lambda:1-\lambda$ where $0 \leq \lambda \leq 1$. Abbreviating notations for simplicity's sake, we define DP-MTL's training objective as follows:

\begin{equation}
L_{DP} = \lambda L_{D} + (1 - \lambda) L_{P}
\label{ldp}
\end{equation}

The objective function of DP-MTL could be thus derived by simply combining the two objective functions.

That is, for all $u, i$ such that $o_{u,i} = o_{i}^{*}$, given the user answer was correct, 
\begin{equation}
\begin{split}
L_{DP}(y_{u,i}, o_{u,i}  ;\boldsymbol{\theta}_u, \textbf{a}_{i}) = \textnormal{log}P(\hat{o}_{u,i} = o_i^*|\boldsymbol{\theta}_u, \textbf{a}_{i}),
\label{coreq}
\end{split}
\end{equation}
and for all $u, i$ such that $o_{i,j} \neq o_{i}^{*}$, given the user answer was incorrect we have:
\begin{equation}
\label{eq:reg}
\begin{split}
L_{DP}(y_{u,i}, o_{u,i}  ;\boldsymbol{\theta}_u, \textbf{a}_{i}) = & \lambda\textnormal{log}[\sum\nolimits_{o_i^j\not=o_i^*}P(\hat{o}_{u,i} = o_i^j|\boldsymbol{\theta}_u, \textbf{a}_{i})]
\\ & +(1-\lambda) \textnormal{log}P(\hat{o}_{u,i}=o_{u,i}|\boldsymbol{\theta}_u, \textbf{a}_{i}).
\end{split}
\end{equation}

Note that the objective function is equivalent to that of Equation \ref{eqn:dirt} when the user chooses the correct option $o_{i}^*$, regardless of $\lambda$ value, as shown in Equation~\ref{coreq}. However, if the user answers incorrectly $o_{u,i} \neq o_{i}^*$ for a given item, not only does the likelihood that the user selects the specific choice increase, but also does the likelihood corresponding to the other incorrect options, proportional to $\lambda$. Figure \ref{fig:learning_dp_mtl} provides a schematic of how DP-MTL controls likelihood for each option.

\begin{figure*}[ht!]
    \centering
    \includegraphics[width=1.8\columnwidth]{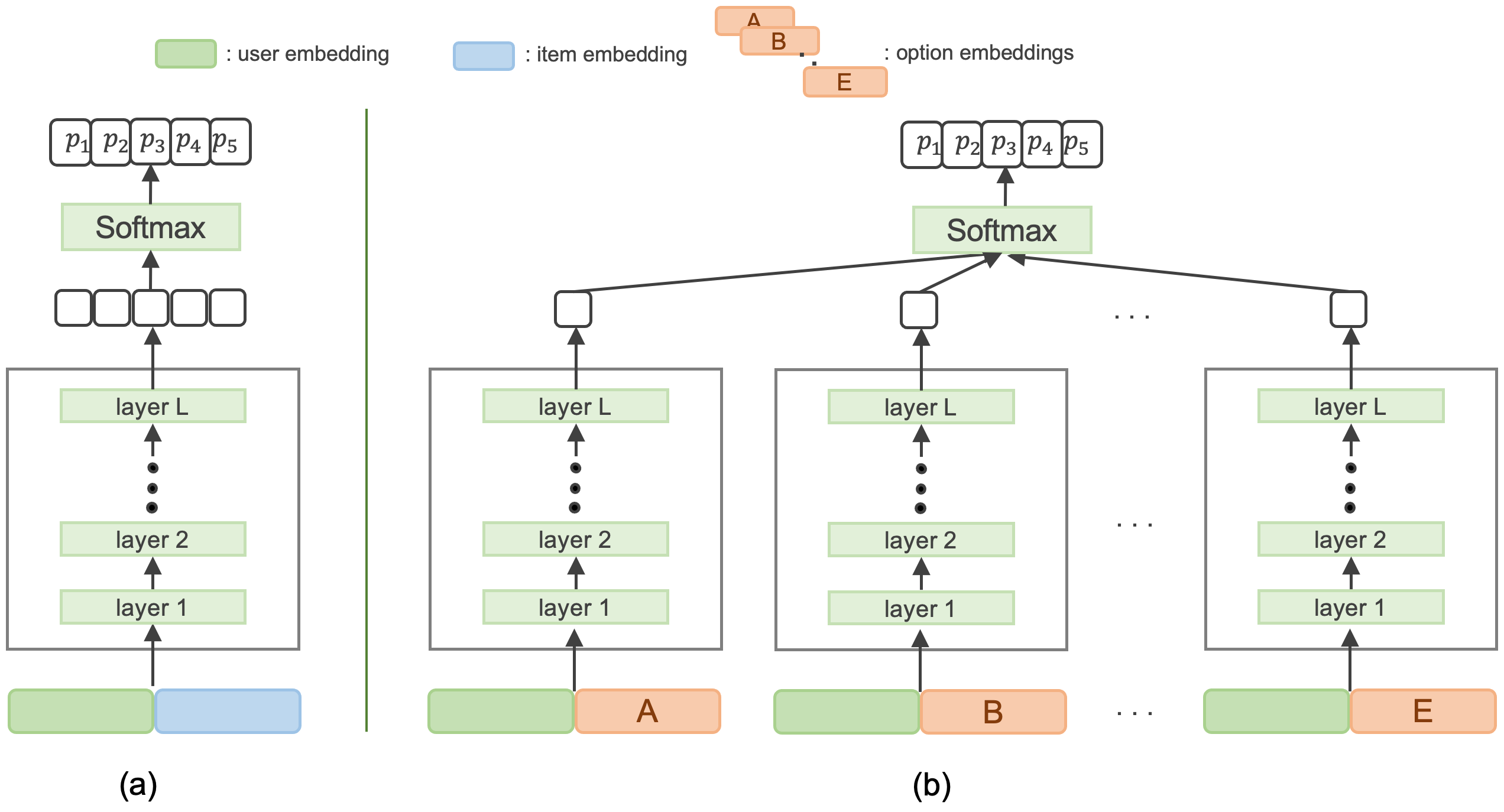}
    \caption{(a) Previous Option Tracing does not use any option information and relies on pre-defined positions for each choice. (b) Our proposed architecture instead uses option embeddings directly to output logits for each option, independent of positions.}
    \label{fig:dp_ncf}
\end{figure*}

Although the underlying motivation is a simple weighted average between the two tasks' losses within a multi-task learning framework, the resulting objective Equation~\ref{eq:reg} shows that the two tasks connect intuitively in a form of regularization. From an option tracing perspective, increasing $\lambda$ from 0 to 1 gradually limits the model's discrimination among the incorrect options. 

As an example where such regularization might help, consider an exam consisting of questions with incorrect option choices that do not discriminate students' skill level to any significant degree (i.e. an exam with low discrimination between student cohorts). Under such circumstances, the optimal $\lambda$ value will be large, and thus the DP-MTL model will be reduced down to a dichotomous option correctness model.

\subsection{Applying DP-MTL to KT models}
In this subsection, we explain how to apply our proposed DP-MTL framework to existing KT models.

\paragraph{DP-IRT}
is the application of DP-MTL to a CF-based option tracing model extended from \citet{vie2019knowledge}.
Using Equation~\ref{ldp} as the objective function, we can combine KT and OT in a straightforward manner.

\paragraph{DP-NMF}
is the application of DP-MTL to NMF~\cite{he2017neural} with some tweaks, as applying DP-MTL to NMF is less trivial than the case of the above DP-IRT.
Recently, \citet{ghosh2021option} proposed to perform OT with NMF. 
However, they treated OT as a multi-class classification problem without giving option information as the input.
For instance, given a set of 5-choice problems, to generate the predictions they share a 5-class Softmax layer as in Figure~\ref{fig:dp_ncf}.
This approach may encounter two issues:
\begin{itemize}
    \item Positional Bias - Prone to learning noises that we do not want to model. For example, information like ``a user may habitually guess for option C'' or ``the real answer was often option B'' can be modeled.
    \item Functionality - Unable to handle cases when the number of choices differ for each question, or when the multiple choice order of each question is mixed for different users.
\end{itemize}
To address these issues, in DP-NMF, we provide option information to the input and generate separate output representations for each option.

\paragraph{DP-BiDKT}
is the application of DP-MTL to DKT~\cite{piech2015deep} with similar tweaks based on the same reasoning as in the above DP-NMF (Figure~\ref{fig:dp_ncf}).
Note that for datasets we consider, we use a Bidirectional LSTM~\cite{lstm}, hence the name BiDKT.
We would also like to point out that we can easily apply similar modifications to Transformer-based~\cite{vaswani2017attention} models \citep{pandey2019self, choi2020towards}.

\section{4. Experiment Setup}

\begin{table*}[ht]
\centering
\begin{tabular}{@{}cccccc@{}}
\toprule
Dataset & Num. Users & Num. Questions & Sparisty & Type          & Description                                                \\ \midrule
ENEM    & 10k        & 185            & 0\%      & Exam     & Brazil's standardized national college entrance exam.                            \\
TOEIC   & 9877       & 13399          & 4\%      & Snapshot & User's 2-week snapshot from \citet{choi2020ednet} \\ \bottomrule
\end{tabular}
\label{tab:datasets}
\caption{A summary of the two datasets used in our experiments: ENEM and TOEIC. }
\end{table*}

\begin{table}[ht]
\centering
\begin{tabular}{@{}ccc@{}}
\toprule
Model                   & Hyper-parameter              & Search Space                \\ \midrule
\multirow{2}{*}{Common} & Mixing Ratio $\lambda$               & $\{0.0, 0.1, ... 1.0\}$           \\
                        & Embedding Dimension          & $\{1, 4, 8, 16, 32, 64\}$         \\ \midrule
NMF                     & \multirow{2}{*}{Num. Layers} & \multirow{2}{*}{$\{1, 2, 3, 4\}$} \\
Bi-DKT                  &                              &                             \\ \bottomrule
\end{tabular}
\label{tab:hyperparams}
\caption{The table shows the hyper-parameter search space for all models and datasets considered in our experiments.}
\end{table}

\label{section:exp}
The proposed DP-MTL framework is evaluated on the three models explained in Section 3 (DP-IRT, DP-NMF, DP-BiDKT) and three tasks (KT, OT, SP), based on two different datasets (ENEM, TOEIC). 
KT and OT serve as primary tasks, while Score Prediction (SP) serves as a downstream task, where we evaluate the quality of the student representation obtained from KT and OT.
For KT, we measure the performance with ROC-AUC (Area under the ROC) as it is a binary classification task, while for OT, we use Accuracy as performance measure. 
For SP, a regression task of predicting the student's exam score, we use Mean Absolute Error (MAE) as performance measure.


In our experiments, we first obtain student and question representations via DP-MTL framework, then fit a simple score prediction model based on training user/student split.
The score prediction module consists of user representation fed into a simple linear regression model followed by isotonic regression. The model's performance is measured by the test MAE metric, which also serves as a quality measure for the student representation $\theta$ from DP-MTL framework.

\subsection{Datasets}

\paragraph{ENEM}
ENEM dataset in our experiment consists of 10000 students' question solving record on 185 questions from 2019 Exame Nacional do Ensino Medio (ENEM) examination. In ENEM, every student solved all 185 questions, thus providing a dense matrix of students and questions. For score prediction task label, we use sum of the 4 section scores for each student\footnote{The entire code, ENEM, and TOEIC datasets are available at: \url{https://github.com/godtn0/DP-MTL}}.


\paragraph{TOEIC}
9877 active users' interaction dataset within an online Intelligent Tutoring System for preparing Test of English for International Communication (TOEIC) exam was used as an additional real-life dataset. The students solved different sets of questions within a question bank of 13399 questions. The score dataset consists of the students' self-reported official TOEIC score out of a total score of 990\footnote{Due to privacy issues, we do not release score prediction dataset for TOEIC.}.

\subsection{Sparsity Ablation}
Due to the large number of questions in TOEIC, the original dataset EdNet~\cite{choi2020ednet} is extremely sparse. For inference performance and sparsity ablation, three different versions of the original dataset is created. Only top N\% of questions (columns) solved by most students and N\% of students (rows) who solved most questions are preserved, where N is set to be 10, 25, and 50. Thus, Top 10\% version yields smallest and most dense (4\% sparsity) interaction matrix, while Top 50\% version yields largest and most sparse (74\% sparsity) one.

Hence, to verify our methodology's robustness against situations with data sparsity, different versions of ENEM training interaction dataset were also created by randomly dropping the student-question interaction pair at different ratios (0\%, 10\%, ..., 70\%). In the following section, we report all three tasks' results based on all imposed sparsity ratios.

\begin{table*}[ht!]
\centering
\setlength{\tabcolsep}{0.8em}
\renewcommand{\arraystretch}{1.4}
\begin{tabular}{cc|ccc|ccc}
\toprule
             &          & \multicolumn{3}{c|}{\textbf{SP-MAE}}                                                             & \multicolumn{3}{c}{\textbf{KT-AUC}}                                                             \\ \hline
\textbf{Dataset}      & \textbf{Sparsity} & \multicolumn{1}{c}{DP-BiDKT} & \multicolumn{1}{c}{DP-IRT} & \multicolumn{1}{c|}{DP-NMF} & \multicolumn{1}{c}{DP-BiDKT} & \multicolumn{1}{c}{DP-IRT} & \multicolumn{1}{c}{DP-NMF} \\ \hline
ENEM    & 0\%      & \textbf{43.1(0.8)}           & 48.2(0.9)                  & 50.2(0.7)                   & 0.7373(0.0)                  & 0.7356(0.9)                & \textbf{0.7381(0.1)}       \\
             & 10\%     & \textbf{52.1(0.8)}           & 60.4(0.7)                  & 62.3(0.6)                   & \textbf{0.7374(0.0)}         & 0.73(0.3)                  & 0.7346(0.4)                \\
             & 20\%     & \textbf{56.6(0.5)}           & 58.6(0.6)                  & 60.9(0.4)                   & \textbf{0.7363(0.5)}         & 0.7324(0.3)                & 0.7345(0.2)                \\
             & 30\%     & \textbf{60.1(0.3)}           & 67.0(0.6)                  & 68.6(0.6)                   & \textbf{0.7291(0.1)}         & 0.7233(0.1)                & 0.7272(0.3)                \\
             & 40\%     & \textbf{71.9(0.4)}           & 75.1(0.4)                  & 77.2(0.6)                   & 0.7213(0.1)                  & \textbf{0.7228(0.1)}       & 0.7178(0.3)                \\
             & 50\%     & \textbf{83.5(0.3)}           & 85.6(0.5)                  & 93.5(0.7)                   & 0.7084(0.0)                  & \textbf{0.7113(0.5)}       & 0.7035(0.3)                \\
             & 60\%     & \textbf{104.0(0.1)}          & 142.0(0.0)                 & 115.5(0.7)                  & \textbf{0.6957(0.0)}         & 0.6647(0.0)                & 0.6934(0.9)                \\
             & 70\%     & \textbf{125.5(0.0)}          & 166.4(0.8)                 & 197.9(0.6)                  & \textbf{0.637(0.1)}          & 0.6192(0.9)                & 0.6161(0.3)                \\ \hline
TOEIC\_Top10 & 4\%     & \textbf{62.2(1.0)}           & 76.7(1.0)                  & 72.9(1.0)                   & \textbf{0.7699(0.1)}         & 0.7661(0.9)                & 0.7481(0.6)                \\
TOEIC\_Top25 & 47\%     & \textbf{58.2(0.9)}           & 69.2(1.0)                  & 69.5(1.0)                   & 0.7809(0.6)                  & \textbf{0.7826(0.9)}       & 0.774(0.9)                 \\
TOEIC\_Top50 & 74\%     & \textbf{59.1(0.6)}           & 69.8(1.0)                  & 69.5(0.6)                   & \textbf{0.849(0.6)}          & 0.7961(0.8)                & 0.7864(0.9) \\
\bottomrule
\end{tabular}
\caption{The table represents the test SP-MAE and KT-AUC for each dataset-model configuration. The figures in the brackets represent the best $\lambda$ value. }
\label{table:all-results}
\end{table*}

\begin{table}[ht]
\small
\centering
\begin{tabular}{p{1cm}p{1cm}|ccc}
\toprule
             &          & \multicolumn{3}{c}{OT-ACC}                                \\
Dataset      & Sparsity & DP-BiDKT             & DP-IRT               & DP-NMF      \\
\midrule
ENEM    & 0\%      & \textbf{0.3851(0.1)} & 0.3842(0.6)          & 0.3818(0.0) \\
             & 10\%     & \textbf{0.389(0.4)}  & 0.3831(0.2)          & 0.3843(0.3) \\
             & 20\%     & \textbf{0.3854(0.4)} & 0.3846(0.2)          & 0.3827(0.2) \\
             & 30\%     & \textbf{0.3824(0.1)} & 0.3815(0.1)          & 0.3789(0.5) \\
             & 40\%     & 0.3769(0.6)          & \textbf{0.3801(0.4)} & 0.375(0.3)  \\
             & 50\%     & 0.3636(0.5)          & \textbf{0.3656(0.7)} & 0.3534(0.0) \\
             & 60\%     & \textbf{0.3455(0.2)} & 0.3422(0.4)          & 0.3392(0.0) \\
             & 70\%     & 0.3004(0.3)          & \textbf{0.3131(0.0)} & 0.279(0.4)  \\
\midrule
TOEIC10 & 4\%     & \textbf{0.6746(0.3)} & 0.6679(0.0)          & 0.6563(0.2) \\
TOEIC25 & 47\%     & \textbf{0.6992(0.5)} & 0.6973(0.1)          & 0.6912(0.2) \\
TOEIC50 & 74\%     & \textbf{0.7421(0.6)} & 0.7144(0.2)          & 0.7142(0.2) \\
\bottomrule
\end{tabular}
\caption{The table represents the test OT-ACC for each dataset-model configuration. The figures in the brackets represent the best $\lambda$ value. }
\label{table:otacc}
\end{table}

\section{5. Results and Discussion}
In order to evaluate the performance of the proposed DP-MTL framework, two sets of evaluation results are reported. First, the performance difference based on different values of $\lambda$ is reported to demonstrate the value of multi-task learning in creating a holistic student representation. Second, the individual performances of individual models on various sparsity levels is reported to determine the most effective MTL model for the tasks of KT and SP. 

\subsection{Impact of DP-MTL: $\lambda$ Ablation}
\begin{figure}[ht!]
    \centering
    \includegraphics[width=1.0\columnwidth]{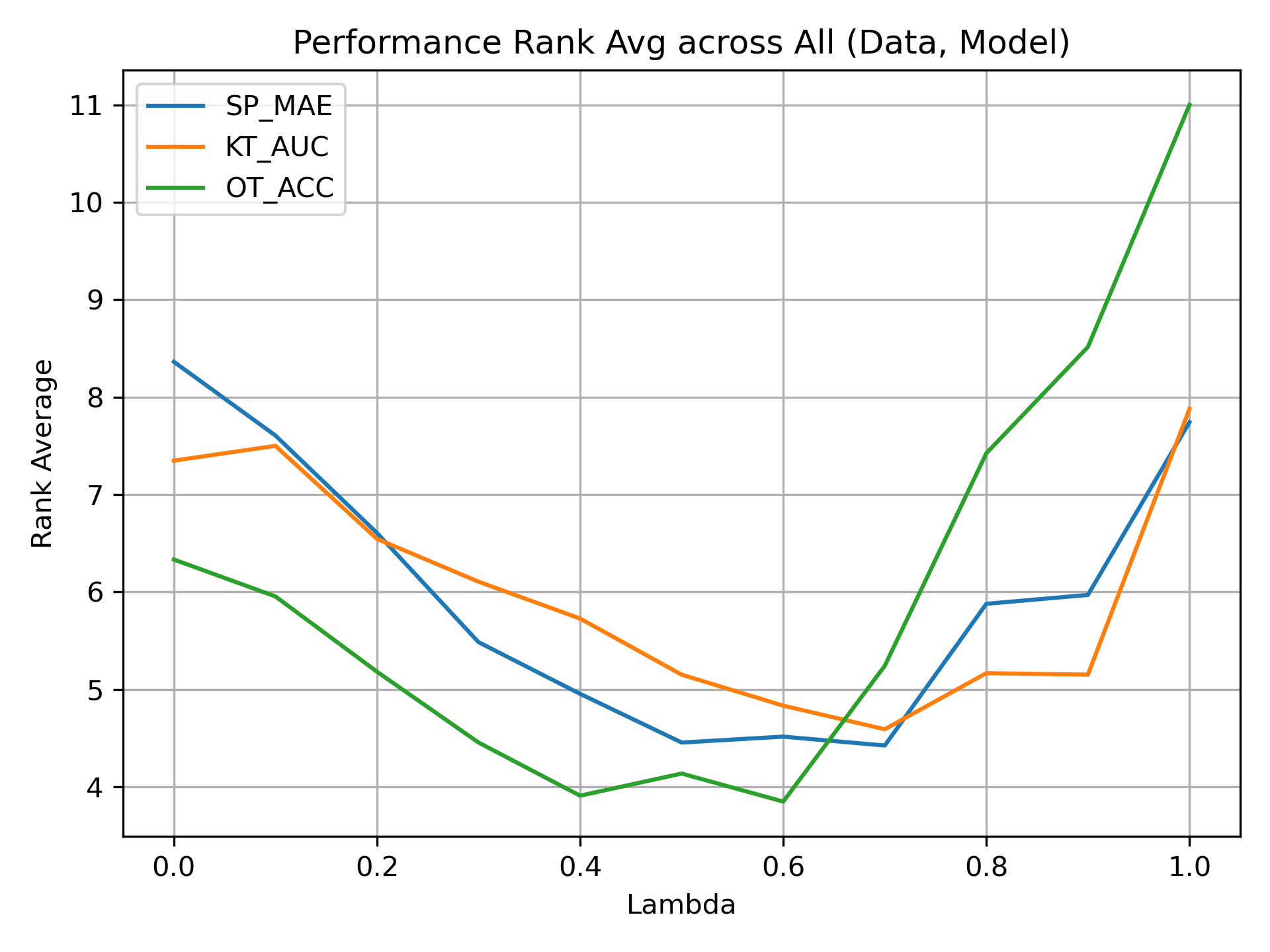}
    \caption{Impact of DP-MTL's $\lambda$ parameter on KT, OT, and SP - y-axis denotes averaged performance rank across all datasets and models in each task. x-axis denotes $\lambda$. Performance rank is a convex curve on the space of $\lambda$. regardless of metrics. }
    \label{fig:task_curves}
\end{figure}

Since different tasks and datasets yield significantly different scales of performance metrics, configuration-wise performance \textbf{rank} of eleven $\lambda$ values (0.0, 0.1, ..., 1.0) were averaged across different datasets and models. The result is shown in Figure \ref{fig:task_curves}. Each line corresponds to different tasks of SP, KT, and OT. Smaller y-axis value of average rank indicates that the performance is relatively superior. For instance, using $\lambda$ value of either 0 or 1 performs significantly worse than $\lambda$ values closer to 0.5, consistently for all three tasks. This convexity serves as a strong empirical evidence of our proposed DP-MTL framework's advantage over the two extreme baseline approaches of KT and OT. We highlight that the multi-task learning of task A (KT) and B (OT) not only improved metrics on the down-stream task C (SP), but also improved the metrics of the original tasks A and B. 

\begin{figure}[ht!]
    \centering
    \includegraphics[width=1.0\columnwidth]{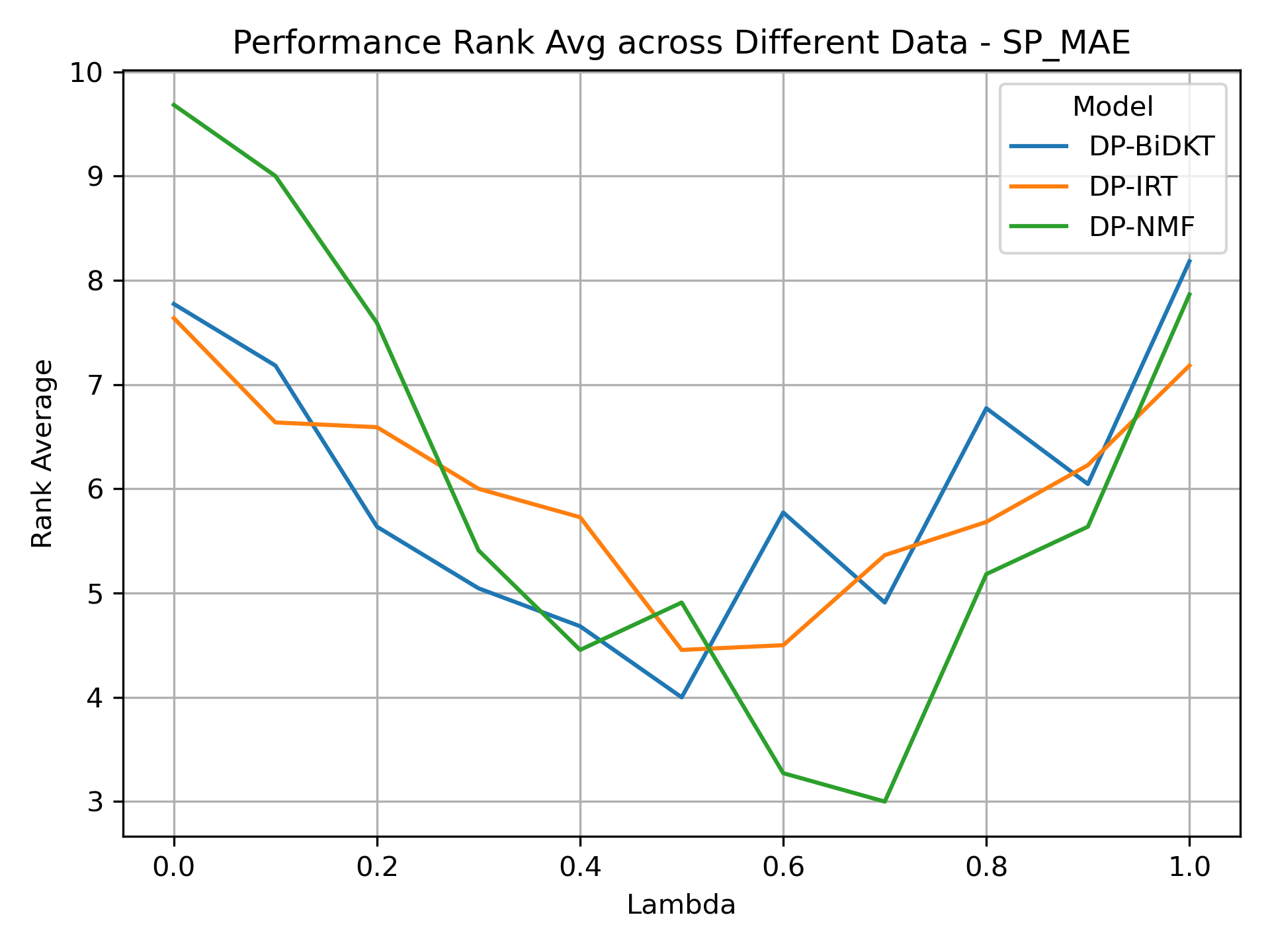}
    \caption{Impact of DP-MTL's $\lambda$ parameter on SP-MAE - y-axis denotes averaged performance rank across all of datasets in SP. x-axis denotes $\lambda$. }
    \label{fig:spmae}
\end{figure}

\paragraph{Score Prediction} All three individual models separately show the desired convex shape of rank average metric with respect to $\lambda$ parameter, as shown in Figure \ref{fig:spmae}. The degree of improvement from KT and OT baselines is largest in DP-NMF model, which has relatively larger number of trainable parameters than the other two models. 

\begin{figure}[ht!]
    \centering
    \includegraphics[width=1.0\columnwidth]{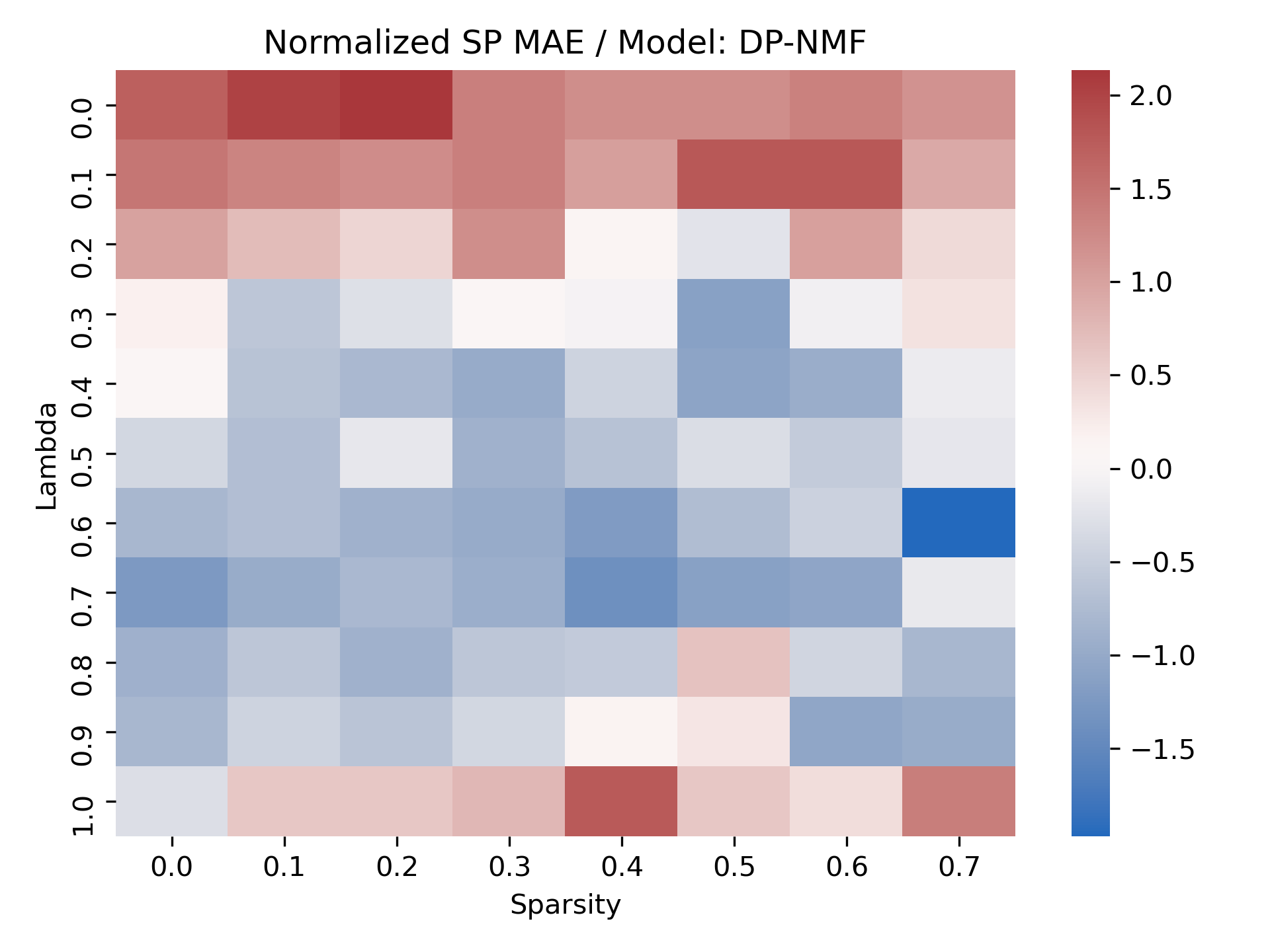}
    \caption{Normalized SP-MAE with DP-NMF in ENEM - Each cell denotes the averaged SP-MAE across all dimensions in conditions with $\lambda$ and sparsity. }
    \label{fig:dpNMF}
\end{figure}

We also note that different models show different trend of optimal $\lambda$ with respect to data sparsity. For DP-NMF, most $\lambda$ hyper-parameters are chosen to be 0.6 and 0.7, consistently. Figure \ref{fig:dpNMF} shows the heatmap of SP-MAE metrics from ENEM dataset standardized within each sparsity ratio setup. Large continuous blue region of smaller MAE emphasizes the advantage from introducing $\lambda$ persists across stable range and across different data sparsity ratios. As opposed to DP-NMF, DP-BiDKT's optimal $\lambda$ value gradually decreases as the dataset sparsity increases. The proposed DP-MTL framework allows the model to tune its attention between KT and OT.

\begin{figure}[ht!]
    \centering
    \includegraphics[width=1.0\columnwidth]{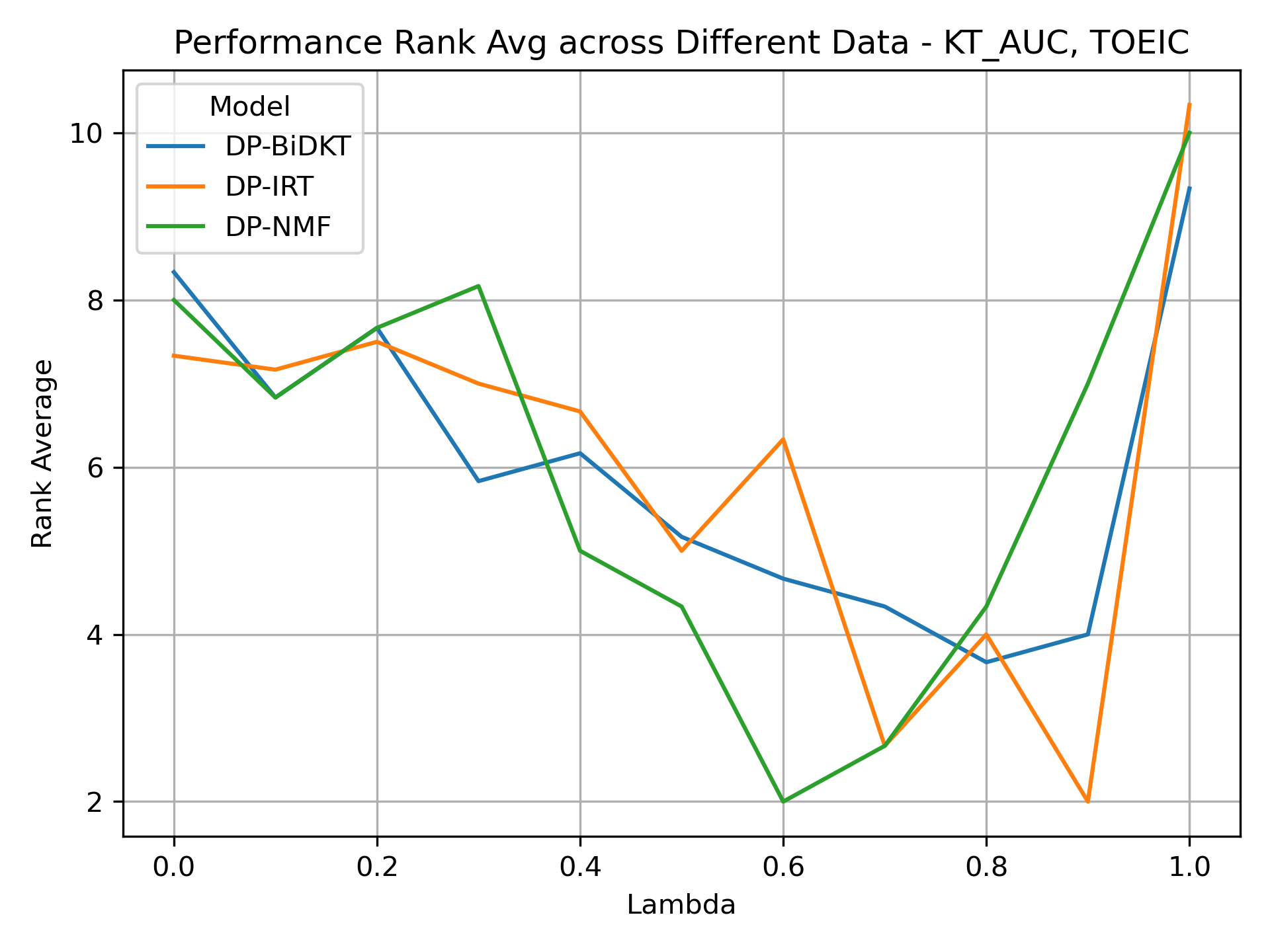}
    \caption{KT-AUC vs $\lambda$, TOEIC}
    \label{fig:ktauc_toeic}
\end{figure}

\paragraph{Knowledge Tracing and Option Tracing}
From KT-AUC block of Table \ref{table:all-results}, most optimal $\lambda$ values in ENEM dataset are closer to 0, as opposed to 1. In other words, tackling OT task alone yielded better results in terms of KT-AUC for ENEM dataset. This trend is particularly strong in TOEIC dataset, as shown in Figure \ref{fig:ktauc_toeic}. For all three models, focusing on KT alone (rank 10) yields worse KT-AUC performance than focusing on OT alone (rank 8). Furthermore, $\lambda$ value between 0.6 and 0.9 leads to sharp improvement of performance(rank 2-4).

\subsection{Model Comparison}
Based on the hyper-parameter configuration chosen from validation set performance, test performance metrics for ENEM and TOEIC dataset's are provided in Table \ref{table:all-results}. First block represents results on Score Prediction-Mean Absolute Error (SP-MAE), and the second block represents Knowledge Tracing-Area Under ROC Curve (KT-AUC). The model entry with best performance is highlighted in bold for each task, and the figures in brackets represent the chosen $\lambda$ parameter in our DP-MTL framework. We reiterate that $\lambda=1$ corresponds to KT/D-IRT scenario, while $\lambda=0$ corresponds to OT/P-IRT scenario.

\paragraph{Score Prediction} We compare the three models in SP task where the assessment is focused on the quality of the extracted student representation. DP-BiDKT significantly outperforms the other models by large margin, consistently across different datasets of different sparsity ratios. Under most sparse conditions, SP-MAE reduction is as high as \textbf{15.3\%} and \textbf{24.9\%} for ENEM and TOEIC dataset, respectively. In general, DP-NMF's capability of fitting into non-linear patterns beyond DP-IRT is not providing any advantage in the SP task. 

\paragraph{Knowledge Tracing and Option Tracing} Although DP-BiDKT model's outperformance is not as significant as that in score prediction task, the model achieved top results in most settings in both knowledge tracing and option tracing task. (Option tracing result is shown in the Appendix Table \ref{table:otacc}.) Also, in the most sparse TOEIC\_Top50 dataset, improvement of DP-BiDKT model over the first-runner-up in both of KT and OT task metrics are \textbf{6.6\%} and \textbf{3.9\%}.

In summary, the empirical results strongly support the efficacy of the proposed DP-MTL framework on all three tasks assessing the quality of user-item representation. The multi-task learning approach for KT and OT not only yielded optimal for the down-stream SP task, but also for KT and OT themselves. Furthermore, our DP-BiDKT architecture achieved significant improvement over standard baseline algorithms by efficient parameter reduction/reusing and novel encoding of user interaction sequence.

\section{6. Conclusion}
This study proposed a multi-task learning framework to include (a) response correctness and (b) the specific response choice of a student to provide a more holistic student assessment model that outperforms the existing single-task baselines. Extensive empirical results from the two datasets and the three tasks (1) showed significant improvement upon existing models (IRT, CF, NMF) and (2) revealed intriguing relationship between KT and OT under various data sparsity conditions. In addition, customized DP-BiDKT architecture was proposed to further improve parameter efficiency and simplify input encoding under our DP-MTL framework, which yielded best performance in most experiment settings. 

Beyond improving KT/OT performance, this work provides an example where better user-item representation can benefit separate down-stream tasks such as student score prediction. Other potential future applications include individualized educational content recommendation and weakness identification based on improved representation learning of students and educational contents.

\bibliography{aaai22}

\end{document}